# An Efficient Secure Distributed Cloud Storage for Append-only Data*

Binanda Sengupta†, Nishant Nikam†, Sushmita Ruj†, Srinivasan Narayanamurthy‡, and Siddhartha Nandi‡

†Indian Statistical Institute, Kolkata
Email: binanda_r@isical.ac.in, nikam_nishant@outlook.com, sush@isical.ac.in

‡NetApp India Pvt. Ltd.
Email: srinivasan.narayanamurthy@netapp.com, siddhartha.nandi@netapp.com

*Abstract*—Cloud computing enables users (clients) to outsource large volume of their data to cloud servers. Secure distributed cloud storage schemes ensure that multiple servers store these data in a reliable and untampered fashion. We propose an idea to construct such a scheme for static data by encoding data blocks (using error-correcting codes) and then attaching authentication information (tags) to these encoded blocks. We identify some challenges while extending this idea to accommodate append-only data. Then, we propose our secure distributed cloud storage scheme for append-only data that addresses the challenges efficiently. The main advantage of our scheme is that it enables the servers to update the parity blocks themselves. Moreover, the client need not download any data (or parity) block to update the tags of the modified parity blocks residing on the servers. Finally, we analyze the security and performance of our scheme.

## I. INTRODUCTION

Cloud service providers offer storage outsourcing facility to their users (clients) who can upload large amount of data to the cloud servers and can later read their data as often as needed. However, as a client (data owner) stores only some metadata for the data file she uploads, the file might get corrupted at the cloud servers, thus making it unavailable at some point of time. Moreover, a malicious cloud server can delete some of the client's data to save space and still claim that it has stored the data properly. *Secure cloud storage* schemes address this problem where the client (or a third party auditor) checks the availability of the file uploaded on the remote server. These schemes typically use concepts of *provable data possession* (PDP) and *proofs of retrievability* (POR) that are introduced by Ateniese et al. [1] and Juels and Kaliski [2], respectively. Such schemes employ an auditing mechanism where the client executes a challenge-response protocol (with the remote server) to check the integrity of her outsourced data. Due to involvement of error-correcting codes (ECCs), POR schemes are less efficient than PDP schemes in general. However, POR schemes preserve the integrity of *all* of the client's data.

Storage servers in practice storage servers are prone to adversarial corruptions and hardware outages [3], [4], [5]. Therefore, the client's data are often distributed to multiple storage servers in order to increase the degree of reliability. Error-correcting codes are used extensively to design distributed storage systems as they provide optimal storage overhead to achieve the same reliability compared to other techniques [6], [7]. Schwarz and Miller [8] exploit algebraic signatures along with error-correcting codes to construct a secure distributed storage, where data blocks in random locations (of the client's data) are checked for integrity. Algebraic signatures can be aggregated into a single signature that reduces the communication overhead significantly.

Bowers et al. [9] propose a secure distributed cloud storage scheme called HAIL (high-availability and integrity layer for cloud storage) that achieves POR guarantees. Encoding of the data blocks of a file is done in two steps: across multiple servers (*dispersal code*) and within each server (*server code*). HAIL enjoys several benefits such as: high reliability, low per-server computation and bandwidth (comparable to single-server POR schemes) and strong adversarial model. Moreover, message authentication codes (MACs) are embedded in the parity blocks — that reduces the storage overhead on the servers. However, HAIL deals with *static* data (that cannot be modified once uploaded to the servers). Extending HAIL for *dynamic* data is left as a future work in [9].

Although generic dynamic data (supporting arbitrary insertions, deletions and modifications) are useful, *append-only* data find numerous applications as well. These primarily include archival data from different sources where data are appended to the existing datasets. For example, data obtained from closed circuit television camera, monetary transactions in banks, medical history of a patient — all must be kept intact with append being the only possible update. Append-only data are also useful for maintaining other log structures (e.g., in certificate transparency schemes [10]). Extending HAIL to support append-only data requires the client to download all the parity blocks making the scheme inefficient (this is discussed in Section VIII).

**Our Contribution** Our contributions in this work are summarized as follows.

- We propose a secure distributed cloud storage scheme for *static* data that borrows the basic storage structure from HAIL. Our scheme offers POR guarantees. We define a security model for distributed cloud storage schemes. Unlike HAIL, an adversary in our scheme cannot modify a dispersal codeword without being detected by the client.

- We extend our scheme for static data to accommodate *append-only* data.
- In our scheme for append-only data, individual servers can update their parity blocks (required for an append) *without any intervention of the client*.
- For an append, the client in our scheme *need not download* any parity (or data) block to recompute the authentication tags corresponding to the (updated) parity blocks. She can only send the relevant changes in these tags to the corresponding servers.
- We use systematic Cauchy Reed-Solomon codes in our scheme for static data and propose a technique to extend such codes to accommodate new symbols appended to the existing message symbols. The corresponding updates on the parity symbols do not touch existing message symbols. The construction of such an extendable code might be useful in other applications as well.
- We provide a prototype implementation of our scheme for append-only data and report the experimental results.

## II. Preliminaries

### A. Notation

Let $\lambda$ be the security parameter. An algorithm $\mathcal{A}(1^\lambda)$ is a probabilistic polynomial-time (PPT) algorithm if its running time is polynomial in $\lambda$ and its output is a random variable which depends on the internal coin tosses of $\mathcal{A}$. An element $a$ chosen from a set $S$ uniformly at random is denoted as $a \xleftarrow{R} S$. A function $f : \mathbb{N} \to \mathbb{R}$ is called negligible in $\lambda$ if $f(\lambda) < \frac{1}{\lambda^c}$ holds for all positive integers $c$ and for all sufficiently large $\lambda$.

### B. Error-Correcting Codes

An $(n, k, d)_\Sigma$-error-correcting code consists of an encoding algorithm Enc: $\Sigma^k \to \Sigma^n$ (encodes a message consisting of $k$ symbols into a longer codeword consisting of $n$ symbols) and a decoding algorithm Dec: $\Sigma^n \to \Sigma^k$ (decodes a codeword to a message), where $\Sigma$ is a finite alphabet and $d$ is the minimum distance (Hamming distance between any two codewords is at least $d$) of the code. The quantity $\frac{k}{n}$ is called the rate of the code. An $(n, k, d)_\Sigma$-error-correcting code can tolerate up to $\lfloor \frac{d-1}{2} \rfloor$ *errors* and $d-1$ *erasures*. If $d = n - k + 1$, we call the code a maximum distance separable (MDS) code. We often specify the parameters of an MDS code by denoting it as an $(n, k)$-MDS code, where $\Sigma$ is implicit and the minimum distance $d = n - k + 1$. Reed-Solomon codes [11] and their extensions are examples of non-trivial linear MDS codes.

Let the message consist of $k$ symbols $m_1, m_2, \ldots, m_k$ from a finite field $\mathbb{F}$ (considered as a column vector of dimension $k$). Then, the classical Reed-Solomon coding employs an $n \times k$ Vandermonde matrix as the *distribution matrix* over $\mathbb{F}$. For encoding, the message vector is multiplied with the distribution matrix to obtain another column vector (codeword) consisting of $n$ symbols $m'_1, m'_2, \ldots, m'_n$ from $\mathbb{F}$ as shown in Fig. 1(a). Moreover, if the code is *systematic*, the first $k$ symbols of the codeword are same as the $k$ symbols of the message. As any $k \times k$ submatrix of the Vandermonde matrix is invertible, the original message can be decoded using any $k$ out of the $n$ symbols of the codeword (for a Reed-Solomon erasure code). Fig. 1(b) shows this decoding procedure, where $i_1, i_2, \ldots, i_k$ are the indices of the $k$ symbols available from the codeword.

$$k \left\{ \begin{bmatrix} a_{11} & a_{12} & \cdots & a_{1k} \\ a_{21} & a_{22} & \cdots & a_{2k} \\ \vdots & \vdots & \ddots & \vdots \\ a_{k1} & a_{k2} & \cdots & a_{kk} \\ \vdots & \vdots & \ddots & \vdots \\ a_{n1} & a_{n2} & \cdots & a_{nk} \end{bmatrix} \right. * \begin{bmatrix} m_1 \\ m_2 \\ \vdots \\ m_k \end{bmatrix} = \begin{bmatrix} m'_1 \\ m'_2 \\ \vdots \\ m'_k \\ \vdots \\ m'_n \end{bmatrix}$$

$(a)$

$$\begin{bmatrix} a_{i_11} & a_{i_12} & \cdots & a_{i_1k} \\ a_{i_21} & a_{i_22} & \cdots & a_{i_2k} \\ \vdots & \vdots & \ddots & \vdots \\ a_{i_k1} & a_{i_k2} & \cdots & a_{i_kk} \end{bmatrix}^{-1} * \begin{bmatrix} m'_{i_1} \\ m'_{i_2} \\ \vdots \\ m'_{i_k} \end{bmatrix} = \begin{bmatrix} m_1 \\ m_2 \\ \vdots \\ m_k \end{bmatrix}$$

$(b)$

Fig. 1: (a) For Reed-Solomon encoding, a codeword of $n \ (= k+s)$ symbols is obtained by multiplying a message vector of dimension $k$ with the $n \times k$ distribution matrix. For a systematic code, $m'_i = m_i$ for all $i \in [1, k]$. (b) For up to $n - k$ erasures, the original message vector is decoded from any $k$ available symbols of the codeword.

## III. Related Work

Ateniese et al. [1] introduce the notion of *provable data possession* (PDP) for static data. In a PDP scheme, the client computes an authentication tag (e.g., message authentication code) for each block of her data file and uploads the file along with the authentication tags. During an audit protocol, the client samples a predefined number of random block-indices and sends them to the server (*challenge* phase). The cardinality of the challenge set is typically taken to be $O(\lambda)$, where $\lambda$ is the security parameter. Depending upon the challenge, the server does some computations over the stored data and sends a proof to the client (*response* phase). Finally, the client checks the integrity of her data based on this proof (*verification* phase). *Almost all* data blocks can be retrieved from a (possibly malicious) server passing an audit with a non-negligible probability. Other PDP schemes (for static or dynamic data) include [12], [13], [14], [15].

Juels and Kaliski [2] introduce *proofs of retrievability* (POR) for static data (Naor and Rothblum [16] give a similar idea for sublinear authenticators). According to Shacham and Waters [17], the retrievability guarantee for *all* data blocks of the outsourced file can be achieved by encoding the original file with an error-correcting code before authenticating (and uploading) the blocks of the encoded file. Due to the redundancy added to the data blocks, the server has to delete or modify a considerable number of blocks to actually delete or modify a data block which makes it difficult for the server to pass an audit. Other POR schemes include [18], [19], [20], [21], [22], [23].

Curtmola et al. [24] and Zhu et al. [25] propose two PDP schemes where the client disseminates her data among multiple servers. Dimakis et al. [26] introduce network coding in the context of distributed storage systems in order to reduce the repair bandwidth. For such a distributed storage system, there are schemes for remote integrity checking [27], [28]. However, network coding is not systematic (i.e., a codeword does not include the input message symbols) that makes these schemes mostly suitable for archival data (with less frequent reads). On the other hand, many distributed storage systems are based on error-correcting codes [6], [8], [9]. Moreover, reads are quite efficient for systematic variants of these codes.

## IV. ERROR-CORRECTING CODES USED IN OUR CONSTRUCTIONS

In our constructions of a secure distributed cloud storage, we use Cauchy Reed-Solomon coding that is a variant of the Reed-Solomon coding. In this section, we briefly discuss Cauchy Reed-Solomon codes over the finite field $\mathbb{Z}_p$ for a prime $p$. Then, we discuss how to extend a Cauchy Reed-Solomon distribution matrix to accommodate new message symbols appended to the end of current symbols.

### A. Cauchy Reed-Solomon Code

For an $(n, k)$-Cauchy Reed-Solomon (CRS) code [29], the distribution matrix is an $n \times k$ matrix $M_{CRS}$ over $\mathbb{Z}_p$, where the submatrix consisting of the first $k$ rows is a $k \times k$ identity matrix and the submatrix consisting of the last $s = n - k$ rows is an $s \times k$ Cauchy matrix. An $s \times k$ Cauchy matrix ($k + s \leqslant p$) is constructed in the following way. Let $X = \{x_1, x_2, \ldots, x_s\}$ and $Y = \{y_1, y_2, \ldots, y_k\}$ be two sets such that the following conditions hold:

- $x_i \in \mathbb{Z}_p$ for all $i \in [1, s]$, and $y_j \in \mathbb{Z}_p$ for all $j \in [1, k]$,
- $X \cap Y = \emptyset$ (thus $\forall i \in [1, s] \; \forall j \in [1, k] \quad x_i - y_j \neq 0$),
- $\forall i \in [1, s] \; \forall l \in [1, s] \backslash \{i\} \quad x_i \neq x_l$,
- $\forall j \in [1, k] \; \forall l \in [1, k] \backslash \{j\} \quad y_j \neq y_l$.

The $s \times k$ Cauchy matrix defined by $X$ and $Y$ consists of the entries $a_{ij} = \frac{1}{x_i - y_j}$, where $i \in [1, s]$, $j \in [1, k]$ (see Fig. 2(a) for an example). The distribution matrix $M_{CRS}$ (see Fig. 2(b)) has the property that any $k \times k$ submatrix is invertible. The encoding and decoding procedures for Cauchy Reed-Solomon (CRS) coding are same as those of the classical Reed-Solomon coding shown in Fig. 1, except that $M_{CRS}$ (instead of the Vandermonde matrix) is used as the distribution matrix.

### B. Extending Cauchy Reed-Solomon Distribution Matrix for Appending Message Symbols

Let an $n \times k$ matrix $M_{CRS}$ defined by $X = \{x_1, x_2, \ldots, x_s\}$ and $Y = \{y_1, y_2, \ldots, y_k\}$ be the distribution matrix for an $(n, k)$-CRS code over $\mathbb{Z}_p$, where the first $k$ rows form a $k \times k$ identity matrix $I_k$ and the last $s = n - k$ rows form an $s \times k$ Cauchy matrix ($n = k + s \leqslant p$). The codeword is obtained by multiplying $M_{CRS}$ with the message vector $\overrightarrow{m} = [m_1, m_2, \ldots, m_k]^T$, where $m_i \in \mathbb{Z}_p$ for $1 \leqslant i \leqslant k$.

Suppose we have to accommodate another symbol $m_{k+1}$ at the end of the message vector $\overrightarrow{m}$. To achieve this, we append another column to the right of $M_{CRS}$. If we want to keep the parameter $n$ unchanged, the number of parity symbols $s$ decreases; that is, we set the parameters $k$ and $s$ with values $k_{new} = k + 1$ and $s_{new} = s - 1$, respectively. However, this reduces the number of parity blocks in the codeword by 1 for every append in $\overrightarrow{m}$ that is not desirable. On the other hand, if we want to keep $s$ unchanged, the number of codeword-symbols $n$ increases; that is, we set $k_{new} = k + 1$ and $n_{new} = n + 1$ with the restriction $n_{new} \leqslant p$. We describe the latter case that we use in our construction.

We choose an element $y \in \mathbb{Z}_p$ such that $y \notin X$ and $y \notin Y$. We set $y_{k_{new}} = y_{k+1} = y$. Then, we construct an $s \times k_{new}$ Cauchy matrix that consists of the entries $a_{ij} = \frac{1}{x_i - y_j}$, where $i \in [1, s]$ and $j \in [1, k_{new}]$. We note that all of the entries in

$$\begin{bmatrix} 8 & 2 & 3 & 4 \\ 7 & 8 & 9 & 3 \\ 6 & 1 & 10 & 5 \end{bmatrix}$$
$$(a)$$

$$\begin{bmatrix} 1 & 0 & 0 & 0 \\ 0 & 1 & 0 & 0 \\ 0 & 0 & 1 & 0 \\ 0 & 0 & 0 & 1 \\ 8 & 2 & 3 & 4 \\ 7 & 8 & 9 & 3 \\ 6 & 1 & 10 & 5 \end{bmatrix}$$
$$(b)$$

$$\begin{bmatrix} 1 & 0 & 0 & 0 & 0 \\ 0 & 1 & 0 & 0 & 0 \\ 0 & 0 & 1 & 0 & 0 \\ 0 & 0 & 0 & 1 & 0 \\ 0 & 0 & 0 & 0 & 1 \\ 8 & 2 & 3 & 4 & 6 \\ 7 & 8 & 9 & 3 & 4 \\ 6 & 1 & 10 & 5 & 7 \end{bmatrix}$$
$$(c)$$

Fig. 2: (a) The $3 \times 4$ Cauchy matrix defined by two sets $X = \{1, 2, 7\}$ and $Y = \{5, 6, 8, 9\}$ over $\mathbb{Z}_{11}$. (b) The corresponding $7 \times 4$ distribution matrix $M_{CRS}$ with the first $k = 4$ rows forming the $4 \times 4$ identity matrix and the last $s = 3$ rows identical with the rows of the Cauchy matrix. (c) The extended $8 \times 5$ $M_{CRS}$ defined by $X = \{1, 2, 7\}$ and $Y = \{5, 6, 8, 9, 10\}$ in order to accommodate the fifth symbol appended to the existing message vector.

this matrix need not be computed afresh. We simply append a column to the previous $s \times k$ Cauchy matrix where $a_{ik_{new}} = \frac{1}{x_i - y_{k_{new}}}$ for every row indexed by $i$. The matrix thus formed is indeed a Cauchy matrix as it satisfies all the conditions mentioned in Section IV-A. Finally, this matrix is appended to the identity matrix $I_{k+1}$ to obtain the updated distribution matrix $M_{CRS}$. An example is shown in Fig. 2(c) where $y$ is chosen to be 10 in $\mathbb{Z}_{11}$ ($k_{new} = 5, n_{new} = 8$).

## V. OVERVIEW OF A SECURE DISTRIBUTED CLOUD STORAGE

In this section, we define a secure distributed cloud storage scheme. The following Section VI describes the security model considered in this work. In Section VII, we provide our basic scheme for static data. In Section VIII, we extend our basic scheme to accommodate append-only data. The distributed cloud storage structures for static data and append-only data are shown in Fig. 3 and Fig. 4, respectively. We assume that there are total $n$ servers the client (data owner) wants to distribute her data file $F$ among. Out of these $n$ servers, she allots $k$ primary servers to store the data blocks of $F$ and $s = n - k$ secondary (parity or redundant) servers to store the parity (or redundant) blocks. Let the servers be denoted by $S_1, S_2, \ldots, S_n$, where the first $k$ servers $S_1, S_2, \ldots, S_k$ are the primary servers and the rest are the secondary servers. Encoding (and decoding) of the data blocks are done *row-wise* (inter-server) and *column-wise* (intra-server) using Cauchy Reed-Solomon (CRS) codes described in Section IV-A and Section IV-B. Authentication tags are attached to some of these blocks in order to detect anomaly when the client audits her data distributed among the servers. After the client initially uploads her file to the servers, the lifetime of the system is split into some time intervals called *epochs*. In each epoch, a set of servers can behave maliciously and corrupt the respective storage (see Section VI for details). We denote by $F_t$ the state of $F$ stored on the servers in the $t$-th epoch.

We define a secure distributed cloud storage scheme for *append-only* data in Definition 1. The security of such a scheme is stated in Definition 2 (see Section VI). A secure distributed cloud storage scheme for *static* data consists of these procedures except the procedure Append.

**Definition 1.** *A secure distributed cloud storage scheme for append-only data consists of the following procedures.*

- Setup($1^\lambda$): *The client generates her secret key $sk$. Let $\mathcal{F}$ be the space of file-identifiers such that each data file is associated with a unique file-identifier from this space.*
- Outsource($F, sk$): *Given a data file $F$, the client chooses a random file-identifier* fid $\stackrel{R}{\leftarrow} \mathcal{F}$ *for $F$. She processes the file $F$ to form another file $F'$ and distributes its shares to the servers. She stores some metadata $d_F$ for the file.*
- Append(fid, $sk, d_F, t$): *The client sends data blocks to be appended at the end of the existing data blocks to the servers and updates her metadata $d_F$. The servers perform append operations at their end.*
- Challenge(fid, $l, d_F, t$): *During the $t$-th epoch, the client generates a random challenge set $Q$ of cardinality $l$ and sends $Q$ to all the servers.*
- Prove($Q, F_t$, fid, $t$): *Given the challenge set $Q$, the servers compute a proof $\Pi$ and send it to the client.*
- Verify($Q, \Pi$, fid, $sk, d_F, t$): *The client outputs 0 if $\Pi$ is a valid proof. Otherwise, the client outputs 1.*
- Redistribute(fid, $sk, d_F, t, F_t, \epsilon_q$): *In the challenge-response phase during the $t$-th epoch, if the client detects that the fraction of corruption exceeds $\epsilon_q$ for some server, the client tries to restore the share of that server. If this process fails, the file is considered to be unavailable.*

## VI. SECURITY MODEL

We assume that a client (cloud user) wants to distribute her data file $F$ among $n$ servers. She chooses $k$ *primary* servers $S_1, S_2, \ldots, S_k$ to store the data blocks of $F$ and $s = n - k$ *secondary* (or redundant) servers $S_{(k+1)}, S_{(k+2)}, \ldots, S_n$ to store the parity (or redundant) blocks. Data blocks are encoded *row-wise* (inter-server or dispersal code) and *column-wise* (intra-server or server code) using CRS codes.

After encoding the data file $F$, the client initially uploads this processed file to the servers. The lifetime of the system is split into some time intervals called *epochs* [9]. A probabilistic polynomial-time (PPT) adversary $\mathcal{A}$ is modeled as *malicious* (i.e., in each epoch it can corrupt *any* number of servers chosen arbitrarily, and it can modify or delete any part of the storage in any server it corrupts). An epoch consists of four phases: an *append* phase (the client appends data blocks to the existing file residing on the servers), a *corruption* phase ($\mathcal{A}$ chooses a set of servers to corrupt), an *audit* phase (the client challenges the servers via spot checking) and a *remediation* phase (the client checks if some corrupted servers provide incorrect responses above a certain threshold fraction $\epsilon_q$). In the remediation phase, if the fraction of corruptions exceeds $\epsilon_q$ for some server, the client reads all the file-shares from each server and tries to decode the original file $F$.

### A. Overview of Security of a Distributed Cloud Storage

A secure distributed cloud storage scheme (for static or append-only data) satisfies the following properties. The formal security definition is given later in this section. The property *freshness* is applicable only for *append-only* data where up-to-date blocks need to be retained by the servers.

1) **Authenticity** The authenticity property requires that the cloud servers cannot produce valid proofs during audits without storing the corresponding blocks untampered, except with a probability negligible in $\lambda$.
2) **Freshness** For append-only data, the client can append new data blocks to the existing data blocks. Moreover, as we will see later in our scheme, appending new data blocks might require updating some of the (parity) blocks. However, a malicious server may discard these changes and keep an old copy of these blocks. Thus, the client must be convinced that the servers have stored the up-to-date blocks.
3) **Retrievability** Retrievability of data requires that, given a PPT adversary $\mathcal{A}$ (possibly corrupting a set of servers) that can respond correctly to a challenge $Q$ with some non-negligible probability, there exists a polynomial-time extractor algorithm $\mathcal{E}$ that can extract *all* data blocks of the file (except with negligible probability) by challenging $\mathcal{A}$ for a polynomial (in $\lambda$) number of times and verifying the responses sent by $\mathcal{A}$. The algorithm $\mathcal{E}$ has a black-box rewinding access to $\mathcal{A}$. *Authenticity* and *freshness* of data restrict the adversary $\mathcal{A}$ to produce valid responses (without storing the authentic and up-to-date data) during these interactions only with some negligible probability.

For append-only data, we describe the security game $\text{GAME}_A$ between the challenger (acting as the client) and the probabilistic polynomial-time adversary $\mathcal{A}$ (acting as the set of corrupt servers) as follows.

- The challenger executes the Setup algorithm to generate the secret key $sk$.
- The adversary $\mathcal{A}$ selects a data file $F$ (associated with a file-identifier fid) to store. The challenger processes $F$ to form another file $F'$ (consisting of data blocks, parity blocks and tags) with the help of $sk$ and distributes the file-shares to the servers. The challenger stores only some metadata $d_F$ to verify the responses given by $\mathcal{A}$ later.

  Each epoch $t = 0, 1, \ldots, q_1$ ($q_1$ is polynomial in the security parameter $\lambda$) consists of the following steps.

  ○ The adversary adaptively chooses a sequence of operations defined by $\{\text{op}_i\}_{1 \leqslant i \leqslant q_2}$ ($q_2$ is polynomial in the security parameter $\lambda$), where $\text{op}_i$ is an authenticated read or an append. Each append request is defined by a message vector $\vec{m}$. For each append, the challenger runs the algorithm Append and stores the latest metadata at her end. The adversary $\mathcal{A}$ can also initiate $q_s$ (polynomial in $\lambda$) audit protocols (challenge-response) with the challenger. During each audit, the challenger challenges $\mathcal{A}$ with a random challenge set $Q$, and the adversary returns a proof to the challenger. The challenger sends the result of the verification to the adversary for each read and each audit (i.e., the output of VerifyRead and the output of Verify).
  ○ The adversary $\mathcal{A}$ can corrupt any number of servers chosen arbitrarily and modify (or delete) any part of the storage in any of these servers.

  In each epoch, the adversary $\mathcal{A}$ can initiate $q_s$ (polynomial in $\lambda$) audit protocols (challenge-response) with the challenger. During each audit, the challenger challenges $\mathcal{A}$ with a random challenge set $Q$, and the adversary returns a proof to the challenger. The challenger verifies the proof and lets $\mathcal{A}$ know the verification result.
- Let $F^*$ be the final state of the file after $\mathcal{A}$ corrupts some of the servers. Now, the challenger challenges the adversary with a random challenge set $Q$, and the adversary returns a proof to the challenger. The challenger verifies the proof and outputs 1 if the proof passes the verification; it outputs 0, otherwise.

$\mathcal{A}$ wins the security game GAME$_A$ if the challenger outputs 1. The security game GAME$_S$ for static data is same as GAME$_A$, except that append queries are not permitted in GAME$_S$.

**Definition 2** (**Security of a Distributed Cloud Storage Scheme**). *A distributed cloud storage scheme for static data (append-only data) is secure if, given any probabilistic polynomial-time adversary $\mathcal{A}$ who can win the security game GAME$_S$ (GAME$_A$) with some non-negligible probability, there exists a polynomial-time extractor algorithm $\mathcal{E}$ that can extract all data blocks of the file by interacting (via challenge-response) with $\mathcal{A}$ polynomially many times.*

### B. Our Security Model vs. HAIL Security Model

We note that our security definition (for static data) is different from that of HAIL [9] in the following ways.
- In HAIL, the security model is designed in order to enable the client to decode (by executing the procedure Redistribute) the original file $F$ from the file-shares of the servers whenever a server provides incorrect responses for a fraction above a threshold $\epsilon_q$. On the other hand, our scheme for static data provides a guarantee that the original file can be retrieved by interacting with a PPT adversary who provides valid proofs with some non-negligible probability during audits. This is why, our security model does not require the procedure Redistribute. However, the client in our scheme can attempt to decode the original file $F$ using the procedure Redistribute in a similar fashion as described in HAIL [9] (see Section VII-B).
- In HAIL, the adversary is *mobile*, and it can corrupt *up to* $b$ servers (chosen arbitrarily) in each epoch, where $b \leqslant \lfloor \frac{n-k}{2} \rfloor$. HAIL is *not* secure according to Definition 2 where the adversary is allowed to corrupt *any* number of servers in an epoch. For example, if the adversary in HAIL could corrupt all of the $n$ servers in a single epoch, it could replace the whole (row-wise) codeword in the second row with that in the first row (which is also a valid codeword). Thus, the adversary could modify the entire second row and still pass the verification during an audit. A sequence of such modifications could make the file $F$ irretrievable at some point of time.

## VII. Secure Distributed Cloud Storage for Static Data

In this section, we construct a secure distributed cloud storage scheme for static data and analyze the security of the scheme. Let $n$ (total number of servers) and $k$ (number of primary servers) be system parameters that are passed as inputs to the procedures of our scheme. The audit phase involves the procedures **Challenge**, **Prove** and **Verify**. We note that the client constructs two distribution matrices $M_{CRS}$ and $M'_{CRS}$ to encode blocks of the data file $F$ *row-wise* and *column-wise*, respectively. We denote the state of $F$ in the $t$-th epoch by $F_t$.

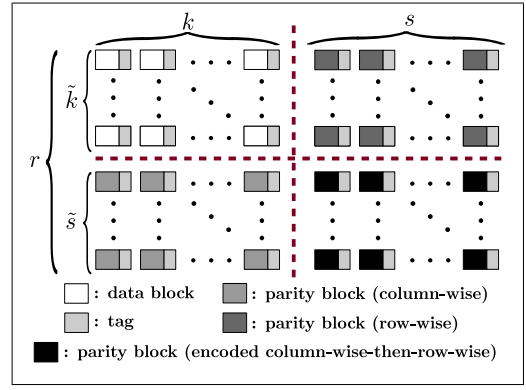

Fig. 3: Distributed storage structure for static data

### A. Our Scheme for Static Data

Our secure distributed cloud storage scheme for static data consists of the following procedures.
- **Setup**($1^\lambda$): The client chooses a random prime $p$ of length $\Theta(\lambda)$ bits. Let $f : \mathcal{K}_{prf} \times \{0,1\}^* \to \mathbb{Z}_p$ be a pseudorandom function (PRF), where $\mathcal{K}_{prf}$ is the key space of the PRF. The client selects an element $\alpha \xleftarrow{R} \mathbb{Z}_p$ and a PRF key $k_{prf} \xleftarrow{R} \mathcal{K}_{prf}$. Her secret key $sk$ is the pair $(\alpha, k_{prf})$. Let $\mathcal{F}$ be the space of file-identifiers.
- **Outsource**($F, sk$): The client chooses a random file-identifier fid $\xleftarrow{R} \mathcal{F}$ for the data file $F$. She divides the file $F$ as $\{m_{ij}\}_{i \in [1, \tilde{k}], j \in [1, k]}$, where each data block $m_{ij} \in \mathbb{Z}_p$ for all $i \in [1, \tilde{k}]$, $j \in [1, k]$. Fig. 3 gives an overview of the storage structure for $F$ distributed among the primary and the secondary servers. For each primary server $S_j$ ($j \in [1, k]$), the client encodes the data blocks $m_{1j}, m_{2j}, \ldots, m_{\tilde{k}j}$ using an $(r, \tilde{k})$ systematic CRS code to form $\tilde{s} = r - \tilde{k}$ parity blocks $m_{(\tilde{k}+1)j}, m_{(\tilde{k}+2)j}, \ldots, m_{rj}$ (with the help of an $r \times \tilde{k}$ matrix $M'_{CRS}$ where $r \leqslant p$). After processing the blocks of the primary servers, the client computes the parity blocks for the secondary servers as follows. For each row $i \in [1, r]$, the client uses an $(n, k)$ systematic CRS code to encode the blocks $m_{i1}, m_{i2}, \ldots, m_{ik}$ into $s = n - k$ parity blocks $m_{i(k+1)}, m_{i(k+2)}, \ldots, m_{in}$ using an $n \times k$ matrix $M_{CRS}$.

The client computes authentication tags (in a similar way as [17]) for the blocks as follows. For each block $m_{ij}$ (where $i \in [1, r]$ and $j \in [1, n]$), she generates a tag

$$\sigma_{ij} = f_{k_{prf}}(i, j) + \alpha m_{ij} \bmod p. \quad (1)$$

Finally, the client sends $\{(m_{ij}, \sigma_{ij})\}_{i \in [1, r]}$ to the $j$-th server $S_j$ for each $j \in [1, n]$.
- **Challenge**(fid, $l, r, t$): During the audit phase in the $t$-th epoch, the client selects $I$, a random $l$-element subset of $[1, r]$ and generates a challenge set $Q = \{(i, \nu_i)\}_{i \in I}$, where each $\nu_i \xleftarrow{R} \mathbb{Z}_p$. Then, the client sends the challenge set $Q$ to all the servers.
- **Prove**($Q, F_t$, fid, $t$): Upon receiving the challenge set $Q = \{(i, \nu_i)\}_{i \in I}$, the $j$-th cloud server $S_j$ computes

$$\mu_j = \sum_{i \in I} \nu_i m_{ij} \bmod p, \quad \sigma_j = \sum_{i \in I} \nu_i \sigma_{ij} \bmod p \quad (2)$$

and sends them to the client (for all $j \in [1, n]$). The responses from all the servers constitute the proof $\Pi$.

- **Verify**$(Q, \Pi, \text{fid}, sk, t)$: Using $Q = \{(i, \nu_i)\}_{i \in I}$ and the proof $\Pi$ sent by the servers, the client checks whether

$$\sigma_j \stackrel{?}{=} \sum_{i \in I} \nu_i f_{k_{prf}}(i, j) + \alpha \mu_j \bmod p \qquad (3)$$

for each $j \in [1, n]$. If any of the equalities does not hold, she outputs 0. Otherwise, the client outputs 1.

- **Redistribute**$(\text{fid}, sk, t, F_t, \epsilon_q)$: During the $t$-th epoch, if the client detects that the fraction (with respect to $|Q|$) of corruption exceeds $\epsilon_q$ for some server, the client reads all the shares of the (possibly corrupted) file from all the servers, tries to recover $F$ (by decoding the file-shares of $F_t$) and distributes new shares to the servers. These distributed file-shares constitute the new state ($F_{t+1}$) of the file. If the decoding procedure fails (that is, the client cannot recover $F$), the file is considered to be unavailable.

### B. Decoding during Redistribution

As we have discussed in Section VI-B, our scheme for static data ensures that the original file $F$ can be retrieved by interacting with a PPT adversary who provides valid proofs with non-negligible probability during audits. However, the client in our scheme can attempt to decode $F$ using the procedure Redistribute in a similar fashion as described in HAIL [9]. We briefly describe this idea of decoding as follows. In order to enable the client to decode the original file, there must be an upper bound on the number of servers $b$ that the adversary can corrupt in each epoch.[1] We use two CRS codes for *erasures* — row-wise coding (using $M_{CRS}$) and column-wise coding (using $M'_{CRS}$). If the client detects, during a redistribution phase, that the fraction of corruption exceeds $\epsilon_q$ for some server, the client downloads all the (possibly corrupted) file-shares from the servers. The decoding procedure works in two steps. In the *first* step, the client executes the decoding procedure row-wise to correct the possible erasures in each row (each of the locations, where the block and its MAC do not match, is marked as an erasure). Therefore, the row-wise decoding can correct up to $s$ erasures. This requirement imposes an upper bound on the value of $b \leqslant n - k = s$ (this bound is $b \leqslant \lfloor \frac{n-k}{2} \rfloor$ for HAIL [9]). If this decoding fails for a particular row (where more than $s$ erasures occur), then the client proceeds to the next step. In the *second* step, the client decodes the column-wise codeword for each primary server (erasures that could not be corrected in the first step). The decoding procedure can recover the original data blocks for a primary server if there are up to $\tilde{s} = r - \tilde{k}$ erasures in the corresponding codeword.

### C. Security

We have described our security model in Section VI, and we analyze the security of our secure distributed cloud storage scheme for static data according to this security model. We state and prove the following theorem in order to analyze the security of our scheme.

**Theorem 1.** *Given that the pseudorandom function (PRF) is secure, our secure distributed cloud storage scheme for static data is secure according to Definition 2.*

*Proof.* We use Claim 1 in order to prove Theorem 1.

**Claim 1.** *Given that the PRF is secure, authenticity of the challenged blocks in each server is guaranteed.*

*Proof.* Authenticity of the challenged blocks demands that, for any server the adversary $\mathcal{A}$ has corrupted, $\mathcal{A}$ cannot produce valid a proof during an audit without correctly storing the challenged blocks (and their respective authentication information) for that server. Without loss of generality, we assume that $\mathcal{A}$ corrupts the $j$-th server $S_j$ ($j \in [1, n]$) along with possibly some other servers, and we prove that Claim 1 still holds for $S_j$. We refer [17] for the detailed proof. Here, we provide a brief sketch of the same.

We assume that, if possible, the adversary $\mathcal{A}$ can produce valid proofs (i.e., they pass the verification) during audits without correctly storing the challenged blocks and their respective authentication information. Without loss of generality, we assume that $\mathcal{A}$ produces a valid (but incorrect) proof $(\sigma'_j, \mu'_j)$ when challenged with $Q = \{(i, \nu_i)\}_{i \in I}$. Let $(\sigma_j, \mu_j) \neq (\sigma'_j, \mu'_j)$ be the correct proof (that is computed honestly) corresponding to the same challenge set $Q = \{(i, \nu_i)\}_{i \in I}$.

As both of the proofs $(\sigma_j, \mu_j)$ and $(\sigma'_j, \mu'_j)$ pass the verification (see Eqn. 3), we have

$$\sigma_j = \sum_{i \in I} \nu_i f_{k_{prf}}(i, j) + \alpha \mu_j \bmod p \qquad (4)$$

and

$$\sigma'_j = \sum_{i \in I} \nu_i f_{k_{prf}}(i, j) + \alpha \mu'_j \bmod p. \qquad (5)$$

We observe, from Eqn. 4 and Eqn. 5, that $\sigma_j = \sigma'_j$ if and only if $\mu_j = \mu'_j$. So, $\sigma_j \neq \sigma'_j$ and $\mu_j \neq \mu'_j$, since $(\sigma_j, \mu_j) \neq (\sigma'_j, \mu'_j)$. If we define $\Delta_\sigma = \sigma'_j - \sigma_j \neq 0$ and $\Delta_\mu = \mu'_j - \mu_j \neq 0$, then we have

$$\Delta_\sigma = \alpha \Delta_\mu. \qquad (6)$$

Therefore, the adversary $\mathcal{A}$ generates a valid (but incorrect) proof exactly when $\Delta_\mu \neq 0$ and Eqn. 6 holds. We note that $\alpha$ is a part of the secret key $sk$ that is not known to $\mathcal{A}$. On the other hand, although $\alpha$ is embedded in each of the authentication tags corresponding to data blocks (see Eqn. 1), it is masked with the output of the pseudorandom function (PRF). As the PRF is *secure*, this masking is computationally indistinguishable from a random masking — which makes the values of the tags independent of $\alpha$. Thus, the value of $\alpha$ is independent of $\mathcal{A}$'s view.

If we consider a sequence of pairs $(\Delta_\sigma, \Delta_\mu)$ computed from $\mathcal{A}$'s responses, then the probability that Eqn. 6 holds for any specific pair (with $\Delta_\mu \neq 0$) in this sequence is $1/p$ (taken over the random choice of $\alpha \stackrel{R}{\leftarrow} \mathbb{Z}_p$). The probability that Eqn. 6 holds for a positive number of pairs is at most $q_s/p$, where $q_s$ is the total number of such interactions initiated by $\mathcal{A}$ (see the security game in Section VI-A). If this bound holds for any fixed sequence of pairs $(\Delta_\sigma, \Delta_\mu)$, it also holds over the random choices of these pairs (as $\alpha$ is independent of $\mathcal{A}$'s view). Now, $q_s$ is polynomial in $\lambda$ and $p = \Theta(2^\lambda)$. So, $\mathcal{A}$ cannot generate (except with negligible probability $q_s/p$) a pair $(\Delta_\sigma, \Delta_\mu)$ such that $\Delta_\mu \neq 0$ and Eqn. 6 holds. Thus, $\mathcal{A}$ cannot produce a valid (but incorrect) proof $(\sigma'_j, \mu'_j) \neq (\sigma_j, \mu_j)$. This completes the proof of Claim 1. □

---

[1]This bound holds only for a successful decoding during a redistribution. On the other hand, the adversary might corrupt *any* number of servers in order to pass an audit (without storing the challenged data). Even then, the adversary cannot cheat with a probability more than negligible in $\lambda$.

Let us denote by $F_j$ the share of the data file $F$ disseminated to the $j$-th server $S_j$ ($j \in [1,n]$). We describe the extraction procedure for (possibly corrupted) $S_j$. The procedure works for other servers as well. We define a polynomial-time extractor algorithm $\mathcal{E}$ that can extract all blocks of $F_j$ (except with negligible probability) by interacting with an adversary $\mathcal{A}$ that wins the security game with some non-negligible probability. As our scheme satisfies the *authenticity* property, the adversary $\mathcal{A}$ cannot produce a valid proof $(\sigma_j, \mu_j)$ for a given challenge set $Q$ without storing the challenged blocks of $F_j$ and their corresponding tags properly, except with some negligible probability (see Claim 1). This means that if the verification procedure outputs 1 during the extraction phase, $\mu_j$ is indeed the correct linear combination of the untampered blocks (that is, $\mu_j = \sum_{i \in I} \nu_i m_{ij} \bmod p$).

Suppose that the extractor algorithm $\mathcal{E}$ wants to extract $l$ blocks (indexed by $J$) of the file $F_j$. It challenges $\mathcal{A}$ with $Q = \{(i, \nu_i)\}_{i \in J}$. If the proof is valid (that is, if the verification procedure outputs 1), $\mathcal{E}$ initializes a matrix $M_{\mathcal{E}}$ with $[\nu_{1i}]_{i \in J}$ as its first row, where $\nu_{1i} = \nu_i$ for each $i \in J$. The extractor challenges $\mathcal{A}$ for the same $J$ but with different random coefficients. If the verification procedure outputs 1 and the vector of coefficients is linearly independent to the existing rows of $M_{\mathcal{E}}$, then $\mathcal{E}$ appends this vector to $M_{\mathcal{E}}$ as a row. The extractor algorithm $\mathcal{E}$ runs this procedure until the matrix $M_{\mathcal{E}}$ has $k$ number of *linearly independent rows*. So, the final form of the *full-rank* matrix $M_{\mathcal{E}}$ is $[\nu_{ui}]_{u \in [1,l], i \in J}$. Consequently, the challenged blocks can be extracted using Gaussian elimination.

Following the way mentioned above, the extractor $\mathcal{E}$ can interact with $\mathcal{A}$ (polynomially many times) in order to extract $\rho$-fraction of blocks (for some $\rho$) present in the file $F_j$ by setting the index set $J$ appropriately. Use of a $\rho$-rate erasure code as the *server code* ensures retrievability of all blocks of $F_j$. This completes the proof of Theorem 1. ∎

### D. Probabilistic Guarantees

Each of the servers is audited with $l$ random locations. Due to the use of a $(r, \tilde{k})$-erasure code as the server code (see Fig. 3), a server has to delete $\tilde{s} = r - \tilde{k}$ blocks (out of total $r$ blocks) in order to actually delete a single block. Then, the server passes an audit with probability $p_{cheat} = (1 - \frac{\tilde{s}}{r})^l = (\frac{\tilde{k}}{r})^{-l}$. The value of $l$ is typically taken to be $O(\lambda)$ in order to make $p_{cheat}$ negligible in $\lambda$.

## VIII. SECURE DISTRIBUTED CLOUD STORAGE FOR APPEND-ONLY DATA

We start with an idea for possible extension of our scheme for static data in order to support append operations and describe some of its challenges. Then, we propose our secure distributed cloud storage for append-only data that addresses these challenges efficiently. We note that extending HAIL [9] for append-only data suffers from two issues stated as follows.

- HAIL uses an adversarial server code [18] that is computationally heavy (due to the use of pseudorandom permutations and encryptions). For each append, parity blocks need to be decrypted (using a secret key of the client), recomputed and permuted again (using another secret key of the client). *This requires the client to download all parity blocks for each server*.
- Parity blocks (for the dispersal code) cannot be extracted from the MACs without knowing the client's secret keys. Thus, even if a classical ECC were used (instead of an adversarial code), the client would have had to download all such parity blocks (in order to update the corresponding MACs) for each append.

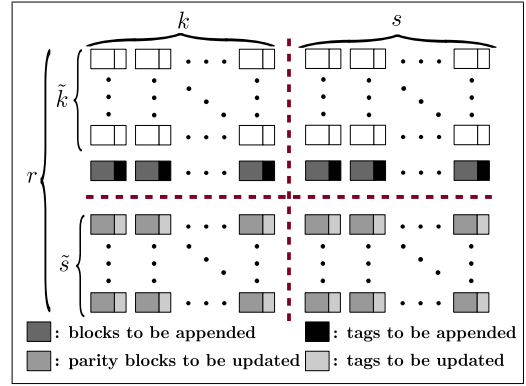

Fig. 4: Distributed storage structure for append-only data

### A. Our Idea for Extension

We assume that the client appends a *row* of data blocks at a time, that is, each server gets a block-tag pair during an append as shown in Fig. 4. The client encodes the new $k$ blocks into $n$ blocks using $M_{CRS}$, generates tags for these blocks and distributes them among the servers. Then, column-wise parity blocks are updated using $M'_{CRS}$ for each server. The client uses two pairs of sets $(X_{row}, Y_{row})$ and $(X_{col}, Y_{col})$ each satisfying the conditions mentioned in Section IV-A. We note that $X_{row}$ and $Y_{row}$ are fixed unless we change the number of primary servers or the number of secondary servers. On the other hand, for column-wise CRS coding using $M'_{CRS}$, the set $Y_{col}$ needs to be changed as discussed in Section IV-B. For each server, the column-wise parity blocks need to be updated depending on the block newly appended to that server.

*1) Challenges:* We discuss about some challenges regarding the proposed idea as follows.
**(a)** For row-wise (or column-wise) CRS coding, the client needs to store the matrices $M_{CRS}$ (or $M'_{CRS}$) to encode data blocks. Alternatively, the client can store only the respective Cauchy matrices at her end to reduce the storage overhead. This overhead can be further alleviated if the client stores only the pairs $(X_{row}, Y_{row})$, $(X_{col}, Y_{col})$ and computes the required entries of the matrices from them on-the-fly. However, storing these pairs also requires $O(p \log p)$ space that is exponential in $\lambda$ as $p = 2^{O(\lambda)}$.

**(b)** A scheme following our basic idea discussed above suffers from the following attack. Let $\tilde{k}$ be the number of data blocks (systematic part) present in the column-wise codeword before the $t$-th append. We fix a row-index $i \in [\tilde{k}+2, r]$ and a column-index $j \in [1, n]$. Let $(m_{ij}, \sigma_{ij})$ and $(m'_{ij}, \sigma'_{ij})$ be the block-tag pairs for the $i$-th block of the $j$-th server $S_j$ *before* and *after* the $t$-th append, respectively. So, we have $\sigma_{ij} = f_{k_{prf}}(i,j) + \alpha m_{ij} \bmod p$ and $\sigma'_{ij} = f_{k_{prf}}(i,j) + \alpha m'_{ij} \bmod p$. Therefore, a (possibly) malicious secondary server $S_j$ can compute $f_{k_{prf}}(i,j)$ and $\alpha$ using $m_{ij}$, $m'_{ij}$, $\sigma_{ij}$ and $\sigma'_{ij}$; and thus it can later generate a valid tag on its $i$-th block.

**(c)** In column-wise CRS coding, for each server, the parity blocks need to be updated depending on the block newly appended to that server. Therefore, for each server, the client needs to download all the parity blocks, update them using the updated $M'_{CRS}$ and upload them (and their corresponding tags) to the corresponding server. This requires a huge client-server communication bandwidth.

*2) Addressing the Challenges*: We describe some remedial measures in order to address the challenges discussed above.

**(a)** We note that the elements in $X_{row}$ and $Y_{row}$ belong to $\mathbb{Z}_p$, where $|X_{row}| = s$ and $|Y_{row}| = k$. We include the first $s$ elements of $\mathbb{Z}_p$ in $X_{row}$ and the last $k$ elements of $\mathbb{Z}_p$ in $Y_{row}$; that is, $X_{row} = \{0, 1, \ldots, s-1\}$ and $Y_{row} = \{p-k, p-k+1, \ldots, p-1\}$. We can easily verify that $X_{row}$ and $Y_{row}$ thus formed indeed satisfy the conditions mentioned in Section IV-A as long as $n = k + s \leq p$. So the knowledge of $i \in [1, s]$ and $j \in [1, k]$ is sufficient to get the entries of $X_{row}, Y_{row}$ and to compute the entries $a_{ij}$ of $M_{CRS}$ on-the-fly. Therefore, the client need not store these sets. $X_{col}$ and $Y_{col}$ can be formed using the same technique, except that $|Y_{col}|$ varies with $\tilde{k}$.

**(b)** The client uses a counter ctr. For the *initial upload*, the client sets the counter ctr to 0 and computes authentication tags

$$\sigma_{ij} = f_{k_{prf}}(i, j, 0) + \alpha m_{ij} \bmod p \quad (7)$$

for all $i \in [1, r]$ and for all $j \in [1, n]$. For each append, the client increments ctr by 1 and updates the tags

$$\sigma'_{ij} = f_{k_{prf}}(i, j, \mathtt{ctr}) + \alpha m'_{ij} \bmod p \quad (8)$$

*only* for $i \in [\tilde{k}+1, r]$ and $j \in [1, n]$, where $m'_{ij}$ (or $\sigma'_{ij}$) is the value of the updated column-wise parity block (or the updated tag for that parity block). We note that, for the first $\tilde{k}$ rows (systematic part), the tags corresponding to blocks (data or parity) in the servers never get updated (as the only operation allowed is append in the $\tilde{k}+1$-th row). Therefore, at any point of time, the value of ctr is 0 for the first $\tilde{k}$ rows and the value of ctr for the rest of the rows is the number of appends that have taken place so far. Let $\sigma_{ij} = f_{k_{prf}}(i, j, \mathtt{ctr}) + \alpha m_{ij} \bmod p$ and $\sigma'_{ij} = f_{k_{prf}}(i, j, \mathtt{ctr}') + \alpha m'_{ij} \bmod p$, where $i \in [\tilde{k}+1, r]$ and $\mathtt{ctr}' > \mathtt{ctr}$. Now, due to the properties of a pseudorandom function, the $j$-th server $S_j$ ($j \in [1, n]$) cannot exploit the knowledge of $\sigma_{ij}$ and $\sigma'_{ij}$ to compute the value(s) of $f_{k_{prf}}(i, j, \mathtt{ctr})$, $f_{k_{prf}}(i, j, \mathtt{ctr}')$ or $\alpha$, for any value of $i \in [\tilde{k}+1, r]$ and for any $\mathtt{ctr}' > \mathtt{ctr}$.

**(c)** For column-wise CRS coding, each server can maintain the updated $M'_{CRS}$ (or it can compute the entries of $M'_{CRS}$ on-the-fly as discussed above) and update its parity blocks using $M'_{CRS}$. *This requires no client-server communication.* Each server $S_j$ ($j \in [1, n]$) updates each of its column-wise parity blocks as follows. Let $\tilde{k}_a$ be the number of data blocks (systematic part) present in the column-wise codeword after the $q$-th append and $m_{\tilde{k}_a j}$ be the newly appended data block for $S_j$. Let $m_{ij}$ and $m'_{ij}$ be the contents of the $i$-th parity block ($i \in [\tilde{k}_a+1, r]$) of $S_j$ before and after the $q$-th append, respectively. The server $S_j$ multiplies $m_{\tilde{k}_a j}$ with the $i$-th entry of the newly added $\tilde{k}_a$-th column of the Cauchy submatrix of $M'_{CRS}$ (we refer to Section IV-B for details). Then, it adds this product to $m_{ij}$ in order to get $m'_{ij}$, the updated content of the parity block. Here, we note that $S_j$ *need not touch any existing data blocks $m_{ij}$ ($i \in [1, \tilde{k}_a-1]$) to update its parity blocks.*

On the other hand, for each server, the authentication tags on the last $\tilde{s} = r - \tilde{k}$ (column-wise parity) blocks need to be updated with the latest value of $i$ and ctr (both are incremented by 1 after an append). We describe the procedure for updating the authentication tag of such a block for $S_j$ ($j \in [1, n]$) as follows. Let $(m_{\tilde{k}_a j}, \sigma_{\tilde{k}_a j})$ be the new block-tag pair to be appended to $S_j$ when the client encodes the $\tilde{k}_a$-th row using row-wise encoding. Let $(m_{ij}, \sigma_{ij})$ and $(m'_{ij}, \sigma'_{ij})$ be the block-tag pairs for the $i$-th block ($i \in [\tilde{k}_a+1, r]$) of $S_j$ before and after the $q$-th append, respectively. Therefore, we have $\sigma_{ij} = f_{k_{prf}}(i, j, q-1) + \alpha m_{ij} \bmod p$ and $\sigma'_{ij} = f_{k_{prf}}(i, j, q) + \alpha m'_{ij} \bmod p$. Let us define $\Delta_\sigma = \sigma'_{ij} - \sigma_{ij} \bmod p$ and $\Delta_m = m'_{ij} - m_{ij} \bmod p$.

As both the row-wise and the column-wise codes used are *linear* codes, it is not hard to see that the content of the $i$-th block of $S_j$ ($i \in [\tilde{k}_a+1, r], j \in [1, n]$) is the same irrespective of whether we get it by column-wise-then-row-wise encoding or by row-wise-then-column-wise encoding. This crucial observation leads us to the fact that $\Delta_m$ can be computed solely from $m_{\tilde{k}_a j}$ and $M'_{CRS}$ as $\Delta_m = m_{\tilde{k}_a j} M'_{CRS}[i, \tilde{k}_a] \bmod p$. Thus, we have

$$\Delta_\sigma = f_{k_{prf}}(i, j, q) - f_{k_{prf}}(i, j, q-1) + \alpha \Delta_m \bmod p. \quad (9)$$

The client sends *only* these $\Delta_\sigma$'s for all relevant parity blocks to the servers, and the servers update the respective authentication tags stored at their end accordingly. Hence, *the client need not download the updated column-wise parity blocks to recompute their authentication tags using Eqn. 8.*

### B. Scheme for Append-only Data

We construct a secure distributed cloud storage for append-only data that addresses the challenges efficiently (see Section VIII-A1 and Section VIII-A2). The procedures **Setup**, **Challenge**, **Prove** and **Redistribute** in our scheme for append-only data are same as those described for *static data* in Section VII-A. We describe the rest of the procedures as follows.

- **Outsource**($F, sk$): The procedure is same as the procedure **Outsource** described in Section VII-A except the following. Instead of using Eqn. 1, the client computes an authentication tag for each block using Eqn. 7 for $i \in [1, r]$ and $j \in [1, n]$. The client and each of the servers initialize the values of ctr stored at their end to be 0.

- **Append**($\mathtt{fid}, sk, \tilde{k}, r, \mathtt{ctr}, t$): Let the client want to append $k$ data blocks to the data file $F$ during the $t$-th epoch. The client increments each of $\tilde{k}$, $r$ and ctr by 1 and updates $M'_{CRS}$. She encodes $k$ data blocks (the row to be appended to $F$) into $n$ blocks using $M_{CRS}$, generates authentication tags

$$\sigma_{\tilde{k} j} = f_{k_{prf}}(\tilde{k}, j, 0) + \alpha m_{\tilde{k} j} \bmod p \quad (10)$$

for all $j \in [1, n]$, and sends the blocks and the tags to the corresponding $n$ servers (see Fig. 4). The servers append the respective blocks (and tags), increment each of $\tilde{k}$, $r$ and ctr they maintain by 1 and update $M'_{CRS}$ at their end. Each of the servers updates its column-wise parity blocks using $M'_{CRS}$ as described in Section VIII-A2. For each server, the client also computes the changes in the existing tags by taking $t = \texttt{ctr}$ and $\tilde{k}_a = \tilde{k}$ in Eqn. 9 and sends them to the servers. The servers update the tags on the existing column-wise parity blocks accordingly.

- **Verify**$(Q, \Pi, \texttt{fid}, sk, \texttt{ctr}, \tilde{k}, t)$: The procedure is same as the procedure **Verify** described in Section VII-A except the following. Instead of using Eqn. 3, the client checks whether

$$\sigma_j \stackrel{?}{=} \sum_{i \in I, i \leqslant \tilde{k}} \nu_i f_{k_{prf}}(i, j, 0) + \sum_{i \in I, i > \tilde{k}} \nu_i f_{k_{prf}}(i, j, \texttt{ctr}) + \alpha \mu_j \bmod p. \quad (11)$$

**Correctness of Verification Eqn. 11** For an honest server $S_j$ ($j \in [1, n]$) storing all the challenged blocks and their corresponding authentication tags correctly, we have

$$\sigma_j = \sum_{i \in I} \nu_i \sigma_{ij} \bmod p \quad \text{[from Eqn. 2]}$$
$$= \sum_{i \in I, i \leqslant \tilde{k}} (\nu_i f_{k_{prf}}(i, j, 0) + \alpha \nu_i m_{ij}) + $$
$$\sum_{i \in I, i > \tilde{k}} (\nu_i f_{k_{prf}}(i, j, \texttt{ctr}) + \alpha \nu_i m_{ij}) \bmod p.$$
$$\text{[from Eqn. 7 and Eqn. 8]}$$
$$= \sum_{i \in I, i \leqslant \tilde{k}} \nu_i f_{k_{prf}}(i, j, 0) + \sum_{i \in I, i > \tilde{k}} \nu_i f_{k_{prf}}(i, j, \texttt{ctr}) + $$
$$\alpha \sum_{i \in I, i \leqslant \tilde{k}} \nu_i m_{ij} + \alpha \sum_{i \in I, i > \tilde{k}} \nu_i m_{ij} \bmod p$$
$$= \sum_{i \in I, i \leqslant \tilde{k}} \nu_i f_{k_{prf}}(i, j, 0) + $$
$$\sum_{i \in I, i > \tilde{k}} \nu_i f_{k_{prf}}(i, j, \texttt{ctr}) + \alpha \mu_j \bmod p. \quad \text{[from Eqn. 2]}$$

Therefore, the proofs provided by an honest server always pass the verification Eqn. 11.

### C. Security

Security proof of our scheme for append-only data is same as that in our scheme for static data (described in Section VII-C), except that the *freshness* (in addition to authenticity and retrievability) of data must be ensured for our secure distributed cloud storage for append-only data (see Section VI-A). The parity blocks in each server are updated for each append. We observe that the first $\tilde{k}$ rows are never updated in our scheme; only the parity rows (i.e., column-wise parity blocks for each server) are updated for an append. If the servers retain the $i$-th ($i \in [\tilde{k}+1, r]$) row of parity blocks with older contents (and tags for an older $\texttt{ctr}$ value), the client can easily detect this anomaly while verifying the proof (using Eqn. 11) as the latest counter value would not match with $\texttt{ctr}$. Thus, the freshness of data is also guaranteed in our scheme.

### D. Number of Parity Blocks in a Column-wise Codeword

In our scheme, we have kept $\tilde{s}$ (the number of parity blocks in a column-wise codeword) fixed throughout a series of append operations in order to avoid inserting rows in the Cauchy submatrix of $M'_{CRS}$ (otherwise, each server has to touch all its data blocks to compute the new parity blocks to be inserted). However, if the value of $\tilde{s}$ becomes very small compared to the increasing value of $\tilde{k}$, then the probability $p_{cheat}$ for a malicious server increases significantly (see Section VII-D). To address this trade-off, we take a parameter $\epsilon_p$ in our system ($0 < \epsilon_p < 1$). If the fraction of parity blocks in a codeword drops below $\epsilon_p$, the client adds some parity blocks to each column-wise codeword to restore the fraction well above the threshold (during Redistribute).

## IX. PERFORMANCE EVALUATION

We have implemented our secure distributed cloud storage scheme for append-only data in order to measure its efficiency. We have run our experiments on a 3.6 GHz Intel Core i5-680 processor. To encode the data blocks of $F$, we use the Jerasure library (version 1.2) [30]. Row-wise CRS encoding uses the distribution matrix $M_{CRS}$. We have run our experiments on total $n = 15$ servers ($k = 9$ primary servers and $s = 6$ secondary servers). Column-wise encoding uses the matrix $M'_{CRS}$. We use (255, 243)-CRS encoding over $\mathbb{GF}[2^8]$ and (65535, 65523)-CRS encoding over $\mathbb{GF}[2^{16}]$. For PRF calculations, we use a cryptographically secure pseudorandom number generator (CPRNG) whose seed is dependent on the index of the respective server and the particular row number of a block (see Eqn. 7, 8, 10 and 11).

$F$ is split into blocks of size 4 KB. Each of the servers contains several disks each of size $2^w \times 4$ KB. After filling a disk, column-wise parity blocks are computed for that disk. We note that an append at the end of previous data blocks affects only the last disk in each server. As arithmetic operations in $\mathbb{Z}_p$ are expensive, we further split each block into chunks each of which is of length $w$ bits ($w = 8$ or $w = 16$). Thus, all the arithmetic operations for row-wise (or column-wise) encoding are done *efficiently* in $\mathbb{GF}[2^w]$. We have run experiments to calculate the time required for different phases in our scheme. We vary parameters such as file size $|F|$ (1–5 GB), query size $|Q|$ (100–1000) and number of failed nodes (1–6). We mention that the reported timing measures also include the time required for disk I/Os and reads (and writes) on the disks.

Table I shows the asymptotic performance of our scheme for append-only data. Fig. 5 depicts the encoding time for varying $|F|$ which includes the total time for row-wise and column-wise encodings. In our experiments, we have taken $\tilde{s} = 12$ for both $w = 8$ and $w = 16$. The encoding time increases for larger $\tilde{s}$ (due to computation of more parity blocks in each server code). Fig. 6 shows the tag generation time for varying $|F|$. We note that, for a fixed file size, the number of chunks in each block for $w = 8$ is greater than that for $w = 16$; and for every chunk in a block, the client computes a PRF value. Thus, the client has to compute more PRF values for $w = 8$.

Fig. 7 and Fig. 8 show the time for proof generation (server side) and proof verification (client side) when the file size increases. Here, we keep the query size constant (500). On the other hand, Fig. 9 and Fig. 10 show the time required for proof generation and proof verification when the size of a query increases. Here, we keep the file size constant (1 GB).

TABLE I: Time needed for different phases in our scheme for append-only data (asymptotic)

| Encoding time (client) | Tag generation time (client) | Proof generation time (server) | Proof verification time (client) | Append time (server) | Client-server communication (for an append) | Reconstruction time (client) |
|---|---|---|---|---|---|---|
| $O(\|F\|)$ | $O(\|F\|)$ | $O(\|Q\|)$ | $O(\|Q\|)$ | $O(\tilde{s})$ | $O(\tilde{s})$ | $O(\|F\|)$ |

Initial encoding of $F$ and generating tags take $O(|F|)$ time as the client has to process all data blocks to obtain parity blocks and to compute their tags. The computations involved in generating a proof and verifying a proof increase linearly as we increase the size of the query $|Q| = |I|$ (see Eqn. 2 and Eqn. 11). As a server has to update its column-wise parity blocks for each append, it takes around $O(\tilde{s})$ time. We note that the client sends the row of blocks to be appended and $\Delta_\sigma$ values for column-wise parity blocks to the servers (see Eqn. 9). So, the client-server communication bandwidth needed for an append is $O(\tilde{s})$. Finally, the reconstruction phase involves decoding $F$ from the file-shares of all the servers — which requires $O(|F|)$ time. Computations involve additions and multiplications in $\mathbb{GF}[2^w]$ and PRF calculations. In our experiments, we take $w = 8$ and $w = 16$ where field operations are efficient.

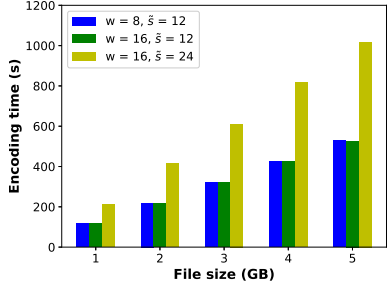

Fig. 5: Encoding time for different $|F|$

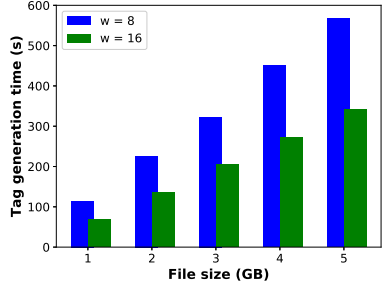

Fig. 6: Tag generation time for different $|F|$

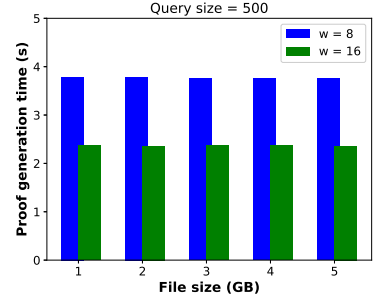

Fig. 7: Proof generation time for different $|F|$

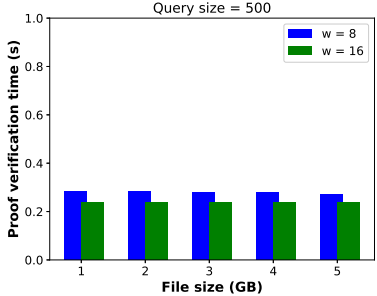

Fig. 8: Proof verification time for different $|F|$

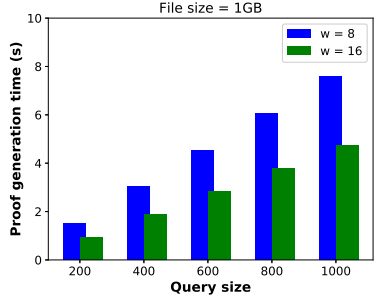

Fig. 9: Proof generation time for different $|Q|$

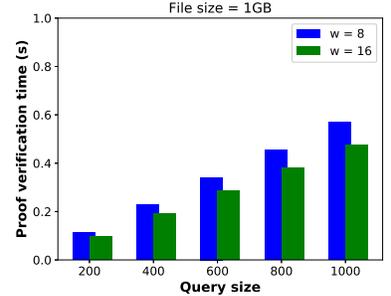

Fig. 10: Proof verification time for different $|Q|$

We note that generating and verifying a proof take $O(|Q|)$ computations and they are independent of $|F|$ (see Table I). As we have mentioned earlier, the client (and the server) has to compute more PRF values for $w = 8$ compared to $w = 16$.

Fig. 11 and Fig. 12 show the time needed for a reconstruction in case one or more servers fail. We can handle at most $s = 6$ server node failures. Fig. 11 depicts the reconstruction time for different $|F|$ when a single node fails. We report the reconstruction time for multiple node failures (for a file of size 1 GB) in Fig. 12. The time for a reconstruction includes the time for decoding and disk I/Os. Fig. 13 shows the time required to append rows to the servers.

## X. Conclusion

In this work, we have proposed a secure distributed cloud storage scheme for static data that achieves POR guarantees. We have presented a technique to extend a systematic Cauchy Reed-Solomon code in order to accommodate new symbols at the end of the existing message symbols. Computations to update the parity symbols do not require reading the existing message symbols. We have used this technique to extend our secure distributed cloud storage scheme (for static data) in order to accommodate append-only data where the client can efficiently append data after the initial data outsourcing. We have also analyzed the security and performance of our scheme.

## Acknowledgment

This project has been made possible in part by a gift from the NetApp University Research Fund, a corporate advised fund of Silicon Valley Community Foundation.

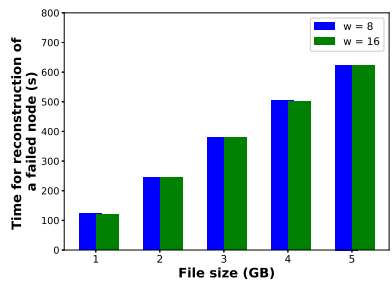

Fig. 11: Time for reconstruction of a failed node for different $|F|$

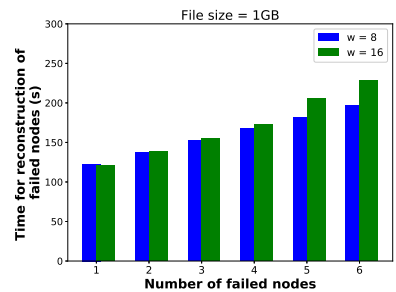

Fig. 12: Time for reconstruction for varying number of failed nodes

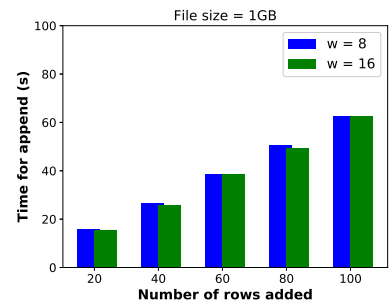

Fig. 13: Time for appending rows